\documentclass[aps,prl,twocolumn,groupedaddress]{revtex4}

\usepackage[dvips]{graphicx,color} 
\usepackage{array,hhline,dcolumn} 
\usepackage{rotating} 

\newcommand{\gtwid}{\mathrel{\raise.3ex\hbox{$>$\kern-.75em\lower1ex\hbox{$\sim$}}}}
\newcommand{\ltwid}{\mathrel{\raise.3ex\hbox{$<$\kern-.75em\lower1ex\hbox{$\sim$}}}}

\begin{document}
%

\title{A Search for Electron Neutrino Appearance at the $\Delta m^2 \sim 1$ eV$^2$ Scale}

\author{A.~A. Aguilar-Arevalo$^{5}$, A.~O.~Bazarko$^{12}$,
        S.~J.~Brice$^{7}$, B.~C.~Brown$^{7}$,
        L.~Bugel$^{5}$, J.~Cao$^{11}$,
        L.~Coney$^{5}$,
        J.~M.~Conrad$^{5}$, D.~C.~Cox$^{8}$, A.~Curioni$^{16}$,
        Z.~Djurcic$^{5}$, D.~A.~Finley$^{7}$, B.~T.~Fleming$^{16}$,
        R.~Ford$^{7}$, F.~G.~Garcia$^{7}$,
        G.~T.~Garvey$^{9}$, C.~Green$^{7,9}$, J.~A.~Green$^{8,9}$, 
        T.~L.~Hart$^{4}$, E.~Hawker$^{15}$,
        R.~Imlay$^{10}$, R.~A. ~Johnson$^{3}$,
        P.~Kasper$^{7}$, T.~Katori$^{8}$, T.~Kobilarcik$^{7}$,
        I.~Kourbanis$^{7}$, S.~Koutsoliotas$^{2}$, E.~M.~Laird$^{12}$,
        J.~M.~Link$^{14}$, Y.~Liu$^{11}$,
        Y.~Liu$^{1}$, W.~C.~Louis$^{9}$,
        K.~B.~M.~Mahn$^{5}$, W.~Marsh$^{7}$, P.~S.~Martin$^{7}$,
G.~McGregor$^{9}$,
        W.~Metcalf$^{10}$, P.~D.~Meyers$^{12}$,
        F.~Mills$^{7}$, G.~B.~Mills$^{9}$,
        J.~Monroe$^{5}$, C.~D.~Moore$^{7}$, R.~H.~Nelson$^{4}$,
        P.~Nienaber$^{13}$, S.~Ouedraogo$^{10}$, R.~B.~Patterson$^{12}$,
        D.~Perevalov$^{1}$, C.~C.~Polly$^{8}$, E.~Prebys$^{7}$,
        J.~L.~Raaf$^{3}$, H.~Ray$^{9}$, B.~P.~Roe$^{11}$,
A.~D.~Russell$^{7}$,
        V.~Sandberg$^{9}$, R.~Schirato$^{9}$,
        D.~Schmitz$^{5}$, M.~H.~Shaevitz$^{5}$, F.~C.~Shoemaker$^{12}$,
        D.~Smith$^{6}$,
        M.~Sorel$^{5}$,
        P.~Spentzouris$^{7}$, I.~Stancu$^{1}$,
        R.~J.~Stefanski$^{7}$, M.~Sung$^{10}$, H.~A.~Tanaka$^{12}$,
        R.~Tayloe$^{8}$, M.~Tzanov$^{4}$,
        R.~Van~de~Water$^{9}$, M.~O.~Wascko$^{10}$, D.~H.~White$^{9}$,
        M.~J.~Wilking$^{4}$, H.~J.~Yang$^{11}$,
        G.~P.~Zeller$^{5}$, E.~D.~Zimmerman$^{4}$ \\
\smallskip
(The MiniBooNE Collaboration) 
\smallskip
}
\smallskip
\smallskip
\affiliation{
$^1$University of Alabama; Tuscaloosa, AL 35487 \\
$^2$Bucknell University; Lewisburg, PA 17837 \\
$^3$University of Cincinnati; Cincinnati, OH 45221\\
$^4$University of Colorado; Boulder, CO 80309 \\
$^5$Columbia University; New York, NY 10027 \\
$^6$Embry Riddle Aeronautical University; Prescott, AZ 86301 \\
$^7$Fermi National Accelerator Laboratory; Batavia, IL 60510 \\
$^8$Indiana University; Bloomington, IN 47405 \\
$^9$Los Alamos National Laboratory; Los Alamos, NM 87545 \\
$^{10}$Louisiana State University; Baton Rouge, LA 70803 \\
$^{11}$University of Michigan; Ann Arbor, MI 48109 \\
$^{12}$Princeton University; Princeton, NJ 08544 \\
$^{13}$Saint Mary's University of Minnesota; Winona, MN 55987 \\
$^{14}$Virginia Polytechnic Institute \& State University; Blacksburg, VA
24061
\\
$^{15}$Western Illinois University; Macomb, IL 61455 \\
$^{16}$Yale University; New Haven, CT 06520\\
}

\date{\today}

\begin{abstract}
The MiniBooNE Collaboration reports first results of a search 
for $\nu_e$ appearance
in a $\nu_\mu$ beam. With two largely independent analyses, we observe no
significant excess 
of events above background for reconstructed neutrino energies above
475 MeV. The data are consistent
with no oscillations within a 
two-neutrino appearance-only oscillation 
model.
\end{abstract}


\maketitle


This Letter reports the initial results from a search for $\nu_\mu
\rightarrow
\nu_e$ oscillations by the MiniBooNE Collaboration.
MiniBooNE was motivated by the result from the Liquid
Scintillator Neutrino Detector (LSND)
experiment \cite{lsnd}, which
has presented evidence for $\bar \nu_\mu \rightarrow \bar
\nu_e$ oscillations at the $\Delta m^2 \sim 1$ eV$^2$ scale. Although 
the Karlsruhe Rutherford Medium Energy Neutrino Experiment (KARMEN) 
observed no evidence for neutrino oscillations 
\cite{karmen}, a joint analysis \cite{joint} showed compatibility at 64\% CL.
Evidence for neutrino oscillations also comes from solar-neutrino
\cite{homestake98,sage99,gallex99,sno,sk02} and
reactor-antineutrino experiments \cite{kamland}, which have
observed $\nu_e$ disappearance at $\Delta m^2 \sim 8\times 10^{-5}$
eV$^2$, and atmospheric-neutrino
\cite{kam,sk98,soudan99,macro01} and long-baseline
accelerator-neutrino
experiments \cite{k2k03,minos06}, which have observed $\nu_\mu$ disappearance
at $\Delta m^2 \sim 3\times 10^{-3}$ eV$^2$.

If all three phenomena are caused by neutrino oscillations, these
three $\Delta m^2$ scales cannot be accommodated in an extension of
the Standard Model that allows only three neutrino mass eigenstates.
An explanation of all three mass scales with neutrino oscillations
requires the addition of one or more sterile neutrinos \cite{sorel04} or
further extensions of the Standard Model ({\it e.g.,} \cite{katori}).

The analysis of the MiniBooNE neutrino data presented here
is performed within a two neutrino appearance-only
$\nu_\mu \rightarrow \nu_e$
oscillation model which uses $\nu_\mu$ events to constrain the predicted
$\nu_e$ rate.  
Other than oscillations between
these two species, we assume no effects beyond the Standard
Model.

The experiment uses the Fermilab Booster
neutrino beam, which is produced from 8 GeV protons
incident on a 71-cm-long by 1-cm-diameter beryllium target.  The proton
beam typically has $4\times 10^{12}$ protons per $\sim 1.6$ $\mu$s beam spill
at a rate of 4 Hz.
The number of protons on target per spill
is measured by two toroids in the
beamline. The target is located
inside a focusing horn, which produces a toroidal magnetic field
that is pulsed in time with the beam at a peak current of 174 kA.
Positively charged pions and
kaons, focused by the horn,
pass through a 60-cm-diameter collimator and can
decay in a 50-m-long tunnel, which is 91 cm in radius
and filled with air at atmospheric pressure.


The center of the detector is 541~m from
the front of the beryllium target and 1.9~m above the center of the 
neutrino beam. 
There is about 3 m of dirt overburden above the
detector, which is a spherical tank of
inner radius 610 cm filled with 800 tons
of pure mineral oil (CH$_2$) with a density of 0.86 g/cm$^3$
and an index of refraction of 1.47.
The light attenuation length in the mineral oil
increases with wavelength from a few cm at 280 nm to over
20 m at 400 nm. Charged particles passing through the oil
can emit both directional Cherenkov light 
and isotropic scintillation light.
An optical barrier separates the detector into two regions, 
an inner volume with a radius of 575~cm and an outer volume 35~cm thick. 
The optical barrier supports 1280 equally-spaced inward-facing 8-inch 
photomultiplier tubes (PMTs), providing 10\% photocathode coverage.
An additional 240 tubes are mounted in the outer 
volume, which acts as a veto shield, detecting particles entering or leaving 
the detector. Two types of PMT are used: 
1198 Hamamatsu model R1408 with 9
stages and 322 Hamamatsu model R5912 with 10 stages.
Approximately 98\% of the PMTs have worked well throughout the data taking
period.

The experiment triggers on every
beam spill, with all PMT hits recorded for a $19.2~\mu$s
window beginning $4.4~\mu$s before the spill.
Other triggers include a random 
trigger for beam-unrelated measurements, a laser-calibration trigger,
cosmic-muon triggers, and a trigger to record neutrino-induced events
from the nearby Neutrinos at the Main Injector (NuMI) beamline \cite{numi}.  
The detector electronics, refurbished from LSND \cite{LSNDNIM},
digitize the times and integrated charges of PMT hits.
PMT hit thresholds are $\sim\! 0.1$ photoelectrons (PE); the
single-PE time resolutions achieved by this system are $\sim\! 1.7$ ns and
$\sim\! 1.2$ ns for the two types of PMTs.
One PE corresponds to $\sim\! 0.2$ MeV of electron energy.
Laser calibration, consisting of optical fibers that run from the laser
to dispersion flasks inside the tank, is run continuously at 3.33 Hz to 
determine PMT gains and time offsets.
Averaged over the entire run, the beam-on livetime of the
experiment is greater than 98\%.

The $\nu_\mu$ energy spectrum peaks at 700 MeV and extends 
to approximately 3000 MeV. Integrated over the neutrino flux, 
interactions in MiniBooNE are mostly charged-current
quasi-elastic (CCQE) scattering (39\%), 
neutral-current (NC) elastic scattering (16\%), 
charged-current (CC) single pion production (29\%), and NC single
pion production (12\%). 
Multi-pion and deep-inelastic scattering
contributions are $<5\%$.
NC elastic scattering, with only a recoil nucleon and a neutrino in
the final state, typically produces relatively little light in the
detector and contributes only 3 events to the final background
estimate.

Table \ref{signal_bkgd} shows the estimated number of events 
with reconstructed
neutrino energy, $E_\nu^{QE}$, between 475 MeV and 1250 MeV after
the complete event selection from all of the significant
backgrounds, where $E_\nu^{QE}$ is determined 
from the reconstructed lepton energy and angle with respect
to the known neutrino direction. The background estimate includes
antineutrino events, which represent $<2\%$ of the total.
Also shown is the estimated number of $\nu_e$
CCQE signal events for the LSND central expectation of
0.26\% $\nu_\mu \rightarrow \nu_e$ transmutation.
Studies of random triggers have established that
no significant backgrounds survive the analysis cuts
other than those due to beam related neutrinos, which can be divided
into either $\nu_\mu$-induced
or $\nu_e$-induced backgrounds.
The small fraction of $\nu_e$ from $\mu$, $K$, and $\pi$ 
decay in the beamline gives
a background that is indistinguishable from oscillations except for the energy
spectrum.
CC $\nu_\mu$ events are distinguished from $\nu_e$ events by the
distinct patterns
of Cherenkov and scintillation light for
muons and electrons, as well as by the observation
of a delayed electron from the muon decay, which is observed $>80\%$
of the time from $\nu_\mu$ CCQE events.
NC $\pi^0$ events with only a single
electromagnetic shower reconstructed are the main $\nu_\mu$-induced
background, followed by radiative $\Delta$ decays giving a single photon,
and then neutrino interactions in the dirt surrounding the 
detector, which can mimic a 
signal if a single photon, mostly from $\pi^0$ decay, 
penetrates the veto and converts in the fiducial volume.
\begin{table}
\caption{\label{signal_bkgd} \em The estimated number of events
with systematic error in the $475<E_\nu^{QE}<1250$ MeV energy range
from all of the significant backgrounds,
together with the estimated number of signal events for
0.26\% $\nu_\mu \rightarrow \nu_e$ transmutation, after the complete 
event selection.}
\begin{ruledtabular}
\begin{tabular}{cc}
Process&Number of Events \\
\hline
$\nu_\mu$ CCQE&$10 \pm 2$ \\
$\nu_\mu e \rightarrow \nu_\mu e$&$7 \pm 2$ \\
Miscellaneous $\nu_\mu$ Events&$13 \pm 5$ \\
NC $\pi^0$&$62 \pm 10$\\
NC $\Delta \rightarrow N \gamma$&$20 \pm 4$ \\
NC Coherent \& Radiative $\gamma$&$<1$ \\
Dirt Events&$17 \pm 3$ \\
\hline
$\nu_e$ from $\mu$ Decay&$132 \pm 10$ \\
$\nu_e$ from $K^+$ Decay&$71 \pm 26$ \\
$\nu_e$ from $K^0_L$ Decay&$23 \pm 7$ \\
$\nu_e$ from $\pi$ Decay&$3 \pm 1$ \\
\hline
Total Background &$358 \pm 35$ \\
\hline
0.26\% $\nu_\mu \rightarrow \nu_e$&$163 \pm 21$ \\
\end{tabular}
\end{ruledtabular}
\end{table}

We use PMT charge and time information in the 19.2 $\mu$s 
window to reconstruct neutrino interactions and identify the product
particles.
This time window is defined as an ``event'' and is
divided into ``subevents'', 
collections of PMT hits clustered in time within $\sim\! 100$ ns.
A $\nu_\mu$ CCQE
event with a muon stopping within the tank may have two subevents:
the first subevent from particles produced at the neutrino interaction, 
the second from the muon decay to an electron.
A $\nu_e$ CCQE event has a single subevent.

To ensure
stable, well-targeted beam at full horn current, it is required that 
the two monitoring toroids agree to within 5\%, the estimated 
transverse containment 
of the beam in the target be greater than 95\%, and the measured horn current 
be within 3\% of its nominal value. The event time at the detector 
must be consistent with the beam delivery time (both determined by GPS), 
and the event must pass a number of data
integrity checks. The beam quality 
requirements reject 0.7\% of the events, while the detector time and 
quality requirements remove a further 1.8\%, with the remaining data
corresponding to $(5.58 \pm 0.12) \times 10^{20}$ protons on target.

Next, events with exactly one subevent (as expected for $\nu_e$ CCQE events)  
are selected. By requiring that the subevent have fewer than 6 hits in
the veto and more than 200 hits in the main tank (above the muon-decay electron 
endpoint), entering cosmic-ray muons and their associated decay electrons
are eliminated. The average time of hits in the subevent is required to be
within the beam time window of 4-7 $\mu$s. These cuts yield a 
cosmic ray rejection of greater than 1000:1.

\begin{figure}[htbp]
\centerline{\includegraphics[height=2.5in]{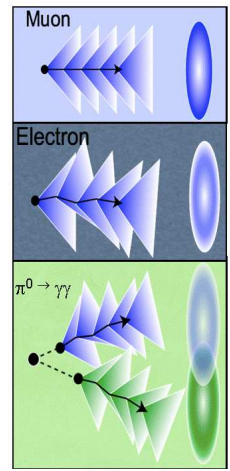}}
\caption{Events in MiniBooNE are reconstructed as either a muon event,
an electron event, or a $\pi^0$ event.}
\label{cartoon}
\end{figure}

After these initial cuts, the surviving events are reconstructed under
four hypotheses: a single electron-like Cherenkov ring, a single
muon-like ring, two photon-like rings with unconstrained kinematics, and
two photon-like rings with $M_{\gamma\gamma}=m_{\pi^0}$
(see Fig. \ref{cartoon}). Photon-like
rings are assumed to be identical to electrons, but allowed to be
independently displaced from the neutrino interaction vertex.  The
reconstruction uses a detailed model of extended-track light production
and propagation in the tank to predict the charge and time of hits on
each PMT.  Event parameters are varied to maximize the likelihood of the
observed hits, yielding the vertex position and time of the event and
the direction, energy, and, for photons, the conversion distance of the
ring(s).
For $\nu_e$ events, the event vertex, direction, and energy are
reconstructed on average with resolutions of 22 cm, $2.8^\circ$, and
11\%, respectively, while NC $\pi^0$ events are reconstructed with a
$\pi^0$ mass resolution of 20 MeV/$c^2$.

The final analysis cuts were designed to isolate a sample of
$\nu_e$-induced events that were primarily CCQE.
The only data that were used in
developing the analysis were samples that Monte Carlo (MC) simulation
had indicated could not contain a significant number of
$\nu_\mu\rightarrow\nu_e$ oscillation events.
We require that the electron-hypothesis event vertex and muon-hypothesis
track endpoint occur at radii $<500$ cm and $<488$ cm, respectively, to
ensure good event reconstruction and efficiency for possible muon decay
electrons.  We require visible energy $E_{vis}>140$ MeV.  We then apply
particle identification (PID) cuts to reject muon and $\pi^0$ events. These
are $E_{vis}$-dependent cuts on $\log(L_e/L_\mu)$,
$\log(L_e/L_{\pi^0})$, and $M_{\gamma\gamma}$, where $L_e$, $L_\mu$, and
$L_{\pi^0}$ are the likelihoods for each event maximized under the electron 
1-ring, muon 1-ring, and fixed-mass 2-ring fits, and
$M_{\gamma\gamma}$ is from the unconstrained two-ring fit.
These also enhance the fraction of CCQE events among the surviving
electron candidates.  Table \ref{data} shows the observed number of $\nu_e$
CCQE candidate events and
the efficiency for $\nu_\mu\rightarrow\nu_e$ CCQE
oscillation events after each cut is applied sequentially. A total
of 380 data events remains after the complete selection.

\begin{table}
\caption{\label{data} \em The observed number of $\nu_e$ CCQE
candidate events and
the efficiency for $\nu_\mu \rightarrow \nu_e$ CCQE oscillation events
after each cut is applied sequentially.}
\begin{ruledtabular}
\begin{tabular}{ccc}
Selection&\#Events&$\nu_e$ CCQE Efficiency \\
\hline
Cosmic Ray Cuts&109,590&$100\%$ \\
Fiducial Volume Cuts&68,143&$55.2 \pm 1.9\%$ \\
PID Cuts&2037&$30.6 \pm 1.4\%$ \\
$475<E^{QE}_\nu<1250$ MeV &380&$20.3 \pm 0.9\%$ \\
\end{tabular}
\end{ruledtabular}
\end{table}

Detailed Monte Carlo simulations of the beam and detector were used
to make initial estimates of the flux and detector efficiencies.
The Booster neutrino beam flux at the detector is
modeled using a GEANT4-based simulation \cite{geant4} of the
beamline geometry.
Pion and kaon production in the target is
parametrized \cite{SW} based on a global fit to proton-beryllium
particle production data \cite{harp}. 
The kaon flux has been cross-checked with high-energy events above 1250
MeV and with an
off-axis muon spectrometer that viewed the secondary beamline from an 
angle of $7^\circ$. This detector determined the flux of muons with 
high transverse momentum, which originate mostly from kaon decays, to 
be consistent with the MC predictions.

The v3 NUANCE \cite{nuance} event generator 
simulates neutrino interactions in mineral 
oil. Modifications are made to NUANCE which include adjustment of the 
axial form factor of the nucleon for quasi-elastic scattering, 
the Pauli blocking model, 
and coherent pion production cross sections 
based on fits 
to MiniBooNE $\nu_\mu$ data \cite{alexis}. 
In addition, the final state interaction 
model has been tuned to reproduce external pion-carbon rescattering data
\cite{ashery}, 
an explicit model of nuclear de-excitation photon emission for carbon has 
been added \cite{deex}, 
and the angular correlations for $\Delta$ decay are modified to
be in accord with the model of Rein and Sehgal \cite{rein_sehgal}.

Particles from NUANCE-generated final
states are propagated through a GEANT3-based simulation \cite{geant3}
of the detector, 
with the subsequent decays, strong, and electroweak interactions
in the detector medium included.
Most particular to MiniBooNE, the emission of
optical and near-UV photons via Cherenkov radiation and
scintillation is simulated, with each photon individually
tracked, undergoing scattering, fluorescence, and reflection,
until it is absorbed \cite{brown}. Small-sample measurements of transmission,
fluorescence, and scattering are used in the model.
Muon decay electrons are used to calibrate both the
light propagation in the detector and the energy
scale. The amount of scintillation light is constrained from
NC elastic scattering events.
The charge and time response of the electronics 
is simulated, and 
from this point onward,
data and MC calculations are treated identically by the
analysis programs.

All of the major $\nu_\mu$-induced backgrounds are constrained by our
measurements outside the signal region. The 
inclusive CC background is verified by comparing data to MC 
calculations for events with
two subevents, where the second subevent has $<200$ tank hits and
is consistent with a muon-decay electron.
As the probability for $\mu^-$
capture in the oil
is $8\%$, there are an order of magnitude
more CC inclusive scattering events with two subevents than with only
one subevent, so that this background is well checked. 
These data events are also modified by moving the hits of the second subevent
earlier in time to model early, inseparable decays which can look more
like an electron.

To determine the NC $\pi^0$ background, $\pi^0$ rates are measured in 
bins of momentum by counting events in the $\gamma\gamma$ mass peak.  
The MC simulation 
is used to correct the production rate for inefficiency, 
background and resolution (corrections are $\sim 10\%$).  
To match the data angular distribution, the 
$\pi^0$ candidates are fit to MC
templates (in mass and angle) for 
resonant and coherent production (generated using the model of Rein and 
Sehgal~\cite{rein_sehgal}) as well as a  template for non-$\pi^0$ background 
events.  The fitted parameters are used to reweight $\pi^0$ from the 
MC calculations
and to constrain the $\Delta\to N\gamma$ rate, which has a branching 
ratio at the peak of the $\Delta$ resonance of 0.56\%.
NC coherent $\gamma$ background \cite{rein} and
NC  radiative $\gamma$ background \cite{goldman}
are both estimated to be negligible. 
The background from interactions in the dirt surrounding the detector
is measured from a sample of inward-pointing 
events inside the tank at high radius.

A sample of $\mathord{\sim}10^5$ candidate $\nu_\mu$ CCQE events 
is obtained by requiring a $\mu$-decay electron with a reconstructed 
vertex consistent with the estimated endpoint of the parent muon's track
(60\% efficiency). 
The observed rate of these $\nu_\mu$ CCQE events is used to correct the 
MC predictions for $\nu_e$ signal events, $\nu_\mu$ CC 
backgrounds, and $\nu_e$ from $\mu$ backgrounds (which share their $\pi$ 
parentage with the $\nu_\mu$ CCQE events).  These constraints increase
the event normalization by 32\% and greatly 
reduce the rate uncertainties on these three components of the final 
analysis sample.

Systematic errors are associated with neutrino fluxes, the detector
model, and neutrino cross sections.
The neutrino flux systematic errors are determined
from the uncertainties of particle production measurements, the
detector model systematic errors are mostly determined from
fits to MiniBooNE data, and the
neutrino cross section systematic errors are determined from MiniBooNE 
data as well as from external sources, both experimental and theoretical. 
These groups of errors are taken to be
independent, and, for each, an individual error matrix is formed
that includes the full correlation among the systematic
parameters.
This is mapped to a matrix describing the correlated errors
in predicted background plus possible signal in eight $\nu_e$
$E_\nu^{QE}$ bins. The final covariance matrix
for all sources of uncertainty (statistical and systematic)
is the sum of the individual error matrices.
The signal extraction is performed by computing the
$\chi^2$ comparing data to predicted background plus a 
($\sin^2(2\theta)$, $\Delta m^2$)-determined contribution from
$\nu_\mu \rightarrow \nu_e$ two-neutrino oscillations in the eight
$E_\nu^{QE}$ bins and minimizing with respect to these two oscillation 
parameters across their physical range.

With the analysis cuts set, 
a signal-blind test of data-MC agreement in 
the signal region was performed.
The full two-neutrino oscillation fit was done in the range $300<E_\nu^{QE}<3000$ MeV and, with no information
on the fit parameters revealed, the sum of predicted background and simulated
best-fit signal was compared to data in several variables, 
returning only the $\chi^2$.
While agreement was good in most of the comparisons, the $E_{vis}$
spectrum had a $\chi^2$ probability of only 1\%.
This triggered further investigation of the backgrounds, focusing on
the lowest energies where $\nu_\mu$-induced backgrounds, 
some of which are difficult to model, are large.  
As part of this study, one more piece of information from the 
signal region was released: unsigned bin-by-bin 
fractional discrepancies in the 
$E_{vis}$ spectrum.  While ambiguous, these reinforced suspicions about
the low-energy region.
Though we found no specific problems with the background estimates, it 
was found that raising the minimum $E_\nu^{QE}$ of the fit region to
475 MeV greatly reduced a number of backgrounds with little impact
on the fit's sensitivity to oscillations.
We thus performed our oscillation fits in the energy range
$475<E_\nu^{QE}<3000$ MeV and opened the full data set.

\begin{figure}[htbp]
\centerline{\includegraphics[height=3.4in]{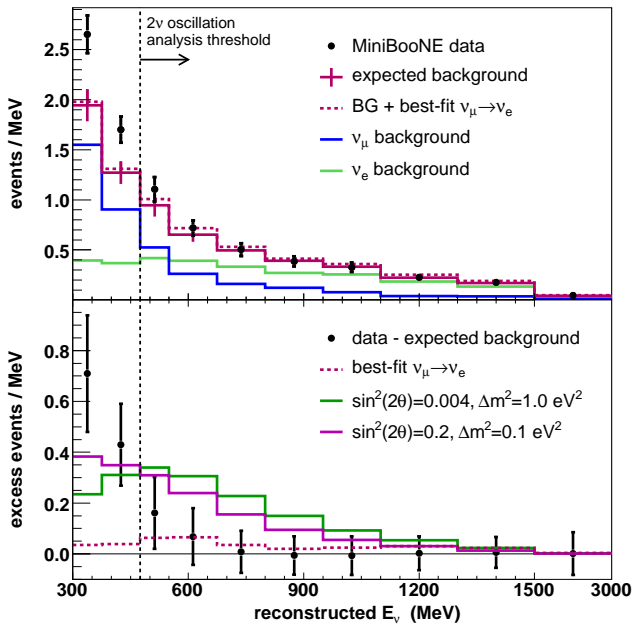}}
\caption{The top plot shows the number of candidate $\nu_e$ events
as a function of $E^{QE}_\nu$.
The points represent the data with statistical error, while the
histogram is the expected background with systematic errors from all sources.
The vertical dashed line indicates the threshold used in the
two-neutrino oscillation analysis. Also shown are the best-fit oscillation
spectrum (dashed histogram) and the background contributions from
$\nu_\mu$ and $\nu_e$ events. The bottom plot shows the number of 
events with the predicted background subtracted
as a function of $E^{QE}_\nu$, where the points represent the data with
total errors and the two histograms correspond to
LSND solutions at high and low $\Delta m^2$.} 
\label{fig:excess}
\end{figure}

The top plot of Fig.~\ref{fig:excess} shows candidate $\nu_e$ events
as a function of $E^{QE}_\nu$.
The vertical dashed line indicates the minimum $E^{QE}_\nu$  
used in the two-neutrino oscillation analysis.
There is 
no significant excess of events ($22 \pm 19 \pm 35$ events) 
for $475<E^{QE}_\nu<1250$ MeV; however, an  
excess of events ($96 \pm 17 \pm 20$ events) is observed below 475 MeV.
This low-energy excess cannot be
explained by a two-neutrino oscillation model, and 
its source is under investigation.
The dashed histogram in Fig.~\ref{fig:excess} shows the predicted spectrum
when the best-fit
two-neutrino oscillation signal is added to the predicted background.  
The bottom panel of the figure shows background-subtracted data with 
the best-fit two-neutrino oscillation and two oscillation points from the 
favored LSND region. The oscillation fit in the $475 < E_\nu^{QE} 
< 3000$ MeV energy range yields a $\chi^2$
probability of 93\% for the null hypothesis, and a probability of 99\% 
for the ($\sin^22\theta = 10^{-3}$, $\Delta m^2 = 4$ eV$^{2}$) best-fit point.

\begin{figure}[htbp]
\centerline{\includegraphics[height=5.0in]{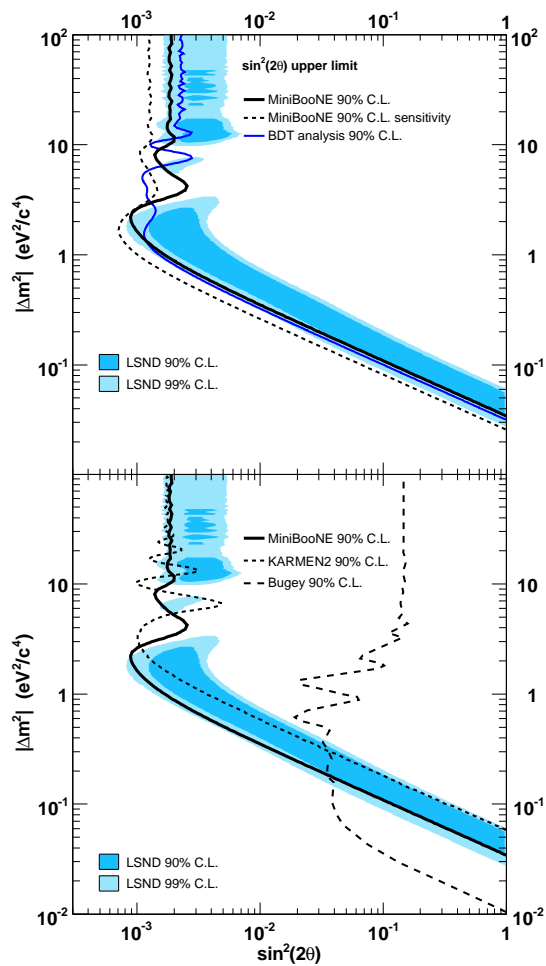}}
\caption{The top plot shows the MiniBooNE 90\% CL limit (thick solid curve) 
and sensitivity (dashed curve) for events with
$475<E_\nu^{QE}<3000$ MeV
within a two neutrino oscillation model. Also shown 
is the limit from the boosted decision tree analysis (thin solid curve)
for events with $300<E_\nu^{QE}<3000$ MeV. The bottom plot shows the
limits from the KARMEN \cite{karmen} and Bugey \cite{bugey} experiments.
The MiniBooNE and Bugey curves
are 1-sided upper limits on $\sin^22\theta$
corresponding to $\Delta \chi^2 = 1.64$, while the KARMEN curve is a 
``unified approach'' 2D contour. The
shaded areas show the 90\% and 99\% CL allowed regions from the LSND
experiment.}
\label{combined_fit}
\end{figure}

A single-sided raster scan
to a two neutrino appearance-only
oscillation model is used in the energy range $475<E_\nu^{QE}<3000$ MeV
to find the 90\% CL limit 
corresponding to $\Delta\chi^2 = \chi^2_{limit} - \chi^2_{best 
fit}= 1.64$.
As shown by the top plot in Fig. \ref{combined_fit}, the
LSND 90\% CL allowed region is excluded at the 90\% CL. 
A joint analysis as a function of $\Delta m^2$, using a 
combined $\chi^2$ of the best fit values and errors for LSND 
and MiniBooNE, excludes at 98\% CL two-neutrino appearance 
oscillations as an explanation of the LSND anomaly. The bottom
plot of Fig. \ref{combined_fit} 
shows limits from the KARMEN \cite{karmen}
and Bugey \cite{bugey} experiments.

A second 
analysis developed simultaneously and with the same blindness
criteria used a different set of reconstruction programs, PID
algorithms, and fitting and normalization processes. The reconstruction
used a simpler model of light emission and propagation.
The PID used 172 quantities such as charge and time likelihoods in
angular bins, $M_{\gamma\gamma}$, and likelihood ratios 
(electron/pion and electron/muon) 
as inputs to boosted decision tree algorithms
\cite{boosting} that 
are trained on sets of simulated signal events and background
events with a cascade-training technique \cite{cascade}.
In order to achieve the maximum sensitivity to oscillations,
the $\nu_\mu$-CCQE data sample with two subevents
were fit simultaneously
with the $\nu_e$-CCQE candidate sample with one subevent.
By forming a $\chi^2$
using both data sets and using the corresponding covariance matrix to relate
the contents of the bins of the two distributions, the errors in the
oscillation parameters that best describe the $\nu_e$-CCQE candidate data
set were well constrained by the
observed $\nu_\mu$-CCQE data.
This procedure is partially equivalent to doing a $\nu_e$ to $\nu_\mu$
ratio analysis where many of the systematic uncertainties cancel.

The two analyses are very complementary, with the
second having a better signal-to-background ratio, but the first having
less sensitivity to systematic errors from detector properties.
These different strengths resulted in very similar oscillation
sensitivities and, when unblinded, yielded the expected overlap
of events and very similar oscillation
fit results. The second analysis also sees
more events than expected at low energy, but with less significance.
Based on the predicted sensitivities before unblinding, 
we decided to present the first
analysis as our oscillation result, with the second as a powerful
cross-check.

In summary, while there is a presently unexplained discrepancy with data lying 
above background
at low energy, there is excellent agreement between data and prediction 
in the oscillation analysis region. If the oscillations of neutrinos
and antineutrinos are the same, this result excludes 
two neutrino 
appearance-only oscillations as an explanation of the LSND anomaly at
98\% CL.

\bigskip

\begin{acknowledgments}
We acknowledge the support of Fermilab, the Department of Energy,
and the National Science Foundation. 
We thank Los
Alamos National Laboratory for LDRD funding.
We acknowledge Bartoszek Engineering for the
design of the focusing horn. 
We acknowledge Dmitri Toptygin, Anna Pla, and Hans-Otto Meyer
for optical measurements of mineral oil.
This research was done using resources provided by the Open Science
Grid, which is supported by the NSF and DOE-SC.
We also acknowledge the use of the LANL Pink cluster
and Condor software in the analysis of the data.
\end{acknowledgments}


\newpage 
\bibliography{prl}

\begin{thebibliography}{10}

\bibitem{lsnd}
C.~Athanassopoulos {\em et~al.}, 
Phys.\ Rev.\ Lett. 75, 2650 (1995);
77, 3082 (1996); 81, 1774 (1998);
A.~Aguilar {\em et~al.},
Phys.\ Rev.\ D 64, 112007 (2001).

\bibitem{karmen}
B.~Armbruster {\em et~al.},
Phys.\ Rev.\ D 65, 112001 (2002).

\bibitem{joint}
E.~Church {\em et~al.},
Phys.\ Rev.\ D 66, 013001 (2002).

\bibitem{homestake98}
B.~T.~Cleveland {\em et~al.},
Astrophys.\ J. 496, 505 (1998).

\bibitem{sage99}
J.~N.~Abdurashitov {\em et~al.}, 
Phys.\ Rev.\ C 60, 055801 (1999).

\bibitem{gallex99}
W.~Hampel {\em et~al.},
Phys.\ Lett.\ B 447, 127 (1999).

\bibitem{sk02}
S.~Fukuda {\em et~al.},
Phys.\ Lett.\ B 539, 179 (2002).

\bibitem{sno}
Q.~R.~Ahmad {\em et~al.},
Phys.\ Rev.\ Lett. 87, 071301 (2001);
Q.~R.~Ahmad {\em et~al.},
Phys.\ Rev.\ Lett. 89, 011301 (2002);
S.~N.~Ahmed {\em et~al.},
Phys.\ Rev.\ Lett. 92, 181301 (2004).

\bibitem{kamland}
K.~Eguchi {\em et~al.},
Phys.\ Rev.\ Lett. 90, 021802 (2003);
T.~Araki {\em et~al.},
Phys.\ Rev.\ Lett. 94, 081801 (2005).

\bibitem{kam}
K.~S.~Hirata {\em et~al.}, 
Phys.\ Lett.\ B 280, 146 (1992);
Y.~Fukuda {\em et~al.},
Phys.\ Lett.\ B 335, 237 (1994).

\bibitem{sk98}
Y.~Fukuda {\em et~al.},
Phys.\ Rev.\ Lett. 81, 1562 (1998).

\bibitem{soudan99}
W.~W. M.~Allison {\em et~al.}, 
Phys.\ Lett.\ B 449, 137 (1999).

\bibitem{macro01}
M.~Ambrosio {\em et~al.},
Phys.\ Lett.\ B 517, 59 (2001).

\bibitem{k2k03}
M.~H.~Ahn {\em et~al.},
Phys.\ Rev.\ Lett. 90, 041801 (2003).

\bibitem{minos06}
D.~G.~Michael {\em et~al.},
Phys.\ Rev.\ Lett. 97, 191801 (2006).

\bibitem{sorel04}
M.~Sorel, J.~M. Conrad, and M.~H.~Shaevitz,
Phys.\ Rev.\ D 70, 073004 (2004).

\bibitem{katori}
T.~Katori, A.~Kostelecky and R.~Tayloe,
Phys.\ Rev.\ D 74, 105009 (2006).





\bibitem{numi}
S.~Kopp,
Phys.\ Rept. 439, 101 (2007).

\bibitem{LSNDNIM}
C.~Athanassopoulos {\em et~al.},
Nucl.\ Instrum.\ Meth. A388, 149 (1997).

\bibitem{geant4}
S.~Agostinelli {\em et~al.},
Nucl.\ Instrum.\ Meth. A506, 250 (2003).

\bibitem{SW}
J.~R.~Sanford and C.~L.~Wang,
Brookhaven National Laboratory, AGS internal reports 11299 and 11479 (1967) 
(unpublished).

\bibitem{harp}
M.~G.~Catanesi {\em et~al.} [HARP Collaboration],
arXiv:hep-ex/0702024;
T.~Abbott {\em et al.}, 
Phys.\ Rev. D45, 3906 (1992);
J.~V.~Allaby {\em et al.}, CERN 70-12 (1970);
D.~Dekkers {\em et al.}, 
Phys.\ Rev. 137, B962 (1965);
G.~J.~Marmer {\em et al.}, 
Phys.\ Rev. 179, 1294 (1969);
T.~Eichten {\em et al.}, 
Nucl.\ Phys. B44, 333 (1972); 
A.~Aleshin {\em et al.}, ITEP-77-80 (1977);
I.~A.~Vorontsov {\em et al.}, ITEP-88-11 (1988).

\bibitem{nuance}
D.~Casper,
Nucl.\ Phys.\ Proc.\ Suppl. 112, 161 (2002).

\bibitem{alexis}
A.~Aguilar-Arevalo {\em et al.}, in preparation.

\bibitem{ashery}
D.~Ashery {\em et~al.},
Phys.\ Rev.\ C 23, 2173 (1981).

\bibitem{deex}
H.~Ejiri, Phys.\ Rev.\ C 48, 1442 (1993);
F.~Ajzenberg-Selove, Nucl.\ Phys.\ A506, 1 (1990);  
G. Garvey private communication.

\bibitem{rein_sehgal}
D.~Rein and L.~M.~Sehgal,
Annals\ Phys. 133, 79 (1981).

\bibitem{geant3}
CERN Program Library Long Writeup W5013 (1993).

\bibitem{brown}
B.~C.~Brown {\em et~al.},
IEEE Nuclear Science Symposium
Conference Record 1, 652 (2004).


\bibitem{rein}
D.~Rein and L.~M.~Sehgal,
Phys.\ Lett.\ B 104, 394 (1981).

\bibitem{goldman}
T.~Goldman, private communication.

\bibitem{bugey}
B.~Achkar {\em et~al.},
Nucl.\ Phys. B434, 503 (1995).


\bibitem{boosting}
B.~P.~Roe {\em et~al.},
Nucl.\ Instrum.\ Meth. A543, 577 (2005);
H.~J.~Yang, B.~P.~Roe, and J.~Zhu,
Nucl.\ Instrum.\ Meth. A555, 370 (2005);
H.~J.~Yang, B.~P.~Roe, and J.~Zhu,
Nucl.\ Instrum.\ Meth. A574, 342 (2007).

\bibitem{cascade}
Y.~Liu and I.~Stancu,
arXiv:Physics/0611267 (to appear in Nucl.\ Instrum.\ Meth.).


\end{thebibliography}

\end{document}